# Evidence of a first-order quantum phase transition of excitons in electron double layers

Biswajit Karmakar[1], Vittorio Pellegrini[1], Aron Pinczuk[2,3], Loren N. Pfeiffer[3] and Ken W. West[3].

[1]NEST INFM-CNR and Scuola Normale Superiore, 56126 Pisa, Italy

[2]Depts of Appl. Phys & Appl. Math. and of Physics, Columbia University, New York, New York 10027, USA

[3]Bell Laboratories, Alcatel-Lucent, Murray Hill, New Jersey 07974, USA

**Complexity in many-particle systems occurs through processes of qualitative differentiation. These are described by concepts such as emerging states with specific symmetries that are linked to order parameters[1,2]. In quantum Hall phases of electrons in semiconductor double layers with large inter-layer electron correlation there is an emergent many body exciton phase with an order parameter that measures the condensate fraction of excitons across the tunneling gap[3,4,5,6,7]. As the inter-layer coupling is reduced by application of an in-plane magnetic field, this excitonic insulating state is brought in competition with a Fermi-metal phase of composite fermions (loosely, electrons with two magnetic flux quanta attached[8,9]) stabilized by intra-layer electron correlation[10]. Here we show that the quantum phase transformation between metallic and excitonic insulating states in the coupled bilayers becomes discontinuous (first-order) by impacts of different terms of the electron-electron interactions that prevail on weak residual disorder. The evidence is based on precise determinations of the excitonic order parameter by inelastic light scattering measurements close to the phase boundary. While there is marked softening of low-lying excitations, our experiments underpin the roles of competing orders linked to quasi-particle correlations in removing the divergence of quantum fluctuations.**

When dominated by quantum-mechanical fluctuations, ground state transformations display a tendency towards quantum criticality associated with collapse of the characteristic energy scale expressed by softening of low-lying excitation modes[2,11,12]. In the presence of competing interactions and weak residual disorder the nature of the quantum phase transition becomes discontinuous as the critical point is approached[13,14,15,16]. One example is offered by the metal-insulator Mott transition which, in the absence of disorder-induced localization effects is believed to be first-order[17]. We show that coupled electron bilayers with strong inter-layer electron correlation realized in double quantum wells in the quantum Hall (QH) regime offer an "ideal" laboratory to study these phenomena.

In these systems the quantum well energy levels split into symmetric and anti-symmetric combinations separated by the tunnelling gap $\Delta_{SAS}$. In the presence of a quantizing perpendicular magnetic field and at total Landau level filling factor $\nu_T = n_T 2\pi l_B^2 = 1$ ($n_T$ is the total electron density in the bilayer and $l_B$ is the magnetic length), the electron bilayer in high-quality AlGaAs/GaAs double quantum wells display two distinct phases that result from the competing impacts of inter- and intra-layer electron interaction[3,4]. These Coulomb interactions are parameterized by $d/l_B$, where $d$ is the distance between the two layers and by $\Delta_{SAS}/E_c$, where $E_c = e^2/\varepsilon l_B$ ($\varepsilon$ is the dielectric constant) is the intra-layer Coulomb energy [3,6]. At sufficiently low $d/l_B$ or large $\Delta_{SAS}/E_c$, the ground states at $\nu_T = 1$ are incompressible QH fluids[3,4]. Early theoretical approaches at $\nu_T = 1$ that considered the role of electron interactions at the mean-field level, described the QH physics of bilayers by assuming full occupation of the lowest symmetric Landau level[18,19]. These studies have linked the disappearing of the incompressible QH phase at low values of $\Delta_{SAS}/E_c$ to the softening of the tunnelling charge excitation at a magneto-roton wavevector, thus suggesting a continuous second-order quantum phase transition (QPT).

Evidence of soft-mode driven QPTs were observed in bilayers at even-integer filling-factor configurations[11,12]. However, evidence of softening of magneto-roton modes at $\nu_T = 1$ is dwindling. Numerical studies and recent inelastic light scattering experiments demonstrated the breakdown of the mean-field picture close to the phase transition highlighting the role of electron correlations at $\nu_T = 1$[5,7]. The light scattering studies have shown that even at sizable $\Delta_{SAS}$ inter-layer correlations favour spontaneous occupation of the excited anti-symmetric level[7]. Such correlated phases, in which anti-symmetric wave-functions are embedded in the ground state, retain the dissipation-less transport characteristic of the quantum Hall regime. The retention of quantum Hall signatures suggests a description of the highly correlated phase at $\nu_T = 1$ linked to emergence of coherent electron-hole excitonic pairs across $\Delta_{SAS}$[20] with a density (the order parameter) that is a fraction of $n_T/2$, as schematically depicted by the cartoon in the top left panel of Fig.1. A different phase occurs for low $\Delta_{SAS}/E_c$ (at sufficiently large $d/l_B$), when inter-layer correlations weaken. This phase appears when intra-layer correlations prevail. The compressible (non-QH) ground state of this phase is currently described as a Fermi metal of composite fermions (CFs) (top-right cartoon in Fig.1).

The keen interest in the interplays between excitonic and CF correlated phases in electron bilayers has prompted consideration of differing scenarios for the states close to the phase boundary and for the discontinuous or continuous character of the transformations[5,21,22,23]. Recent experimental studies have stressed possible roles of electron spin [24,25]. Investigations of the character of the QPTs in bilayers, however, have been hindered by the uncertain role of disorder, that could possibly result to quasi-continuous changes due to smearing of the free-energy landscape and presence of domains of the two phases close to the phase boundary[21,26,27].

To determine the character of the QPT in bilayers, we employ the technique of resonant inelastic light scattering and study the low-lying spin excitations in the excitonic and CF phases, as shown in the middle panels of Fig.1 (description of sample and inelastic light scattering set-up and selection rules are reported in the Methods). We developed an in-situ rotational stage that allows tilting the bilayer sample with respect to the total magnetic field direction and retains the capability to perform optical experiments at temperatures below 100mK under light illumination. The in-plane magnetic field reduces the tunnelling gap according to the relation[28] $\Delta_{SAS}(\theta) = \Delta_{SAS}(\theta=0)\exp(-(\tan(\theta)d/2l_B)^2)$ with a precision of $0.1°$ (precision on $\Delta_{SAS}(\theta)/E_c$ of $1 \times 10^{-4}$). Our method allows to selectively investigating the properties of the two competing phases despite phase coexistence. This together with the precision achieved in the reduction of $\Delta_{SAS}/E_c$ enables the investigation of the evolution of the correlated phases with unprecedented accuracy.

The left-bottom panel of Fig.1 reports a representative example of a spin excitation spectrum in the excitonic phase while spin spectra at different values of $\theta$ (i.e. different values of $\Delta_{SAS}(\theta)/E_c$) are reported in Fig.2. The spin-flip tunnelling mode (SF$_{SAS}$) appears at energies above the spin-wave (SW) excitation at the 'bare' Zeeman energy ($E_z=g\mu_B B_T$ where $g=0.4$ is the gyromagnetic factor, $\mu_B$ the Bohr magnetron and $B_T$ the total magnetic field) by the Larmor theorem. Signatures of the excitonic phase are manifested in the intensity and energy position of SF$_{SAS}$: the intensity of this excitation is linked to the spatial extent of the excitonic domains and vanishes at the transition to the CF metallic state. The energy also carries key information on the excitonic phase. In fact, within Hartree-Fock framework, the SF$_{SAS}$–SW energy splitting $\delta(\theta)$ is equal to $\Delta_{SAS}(\theta)$. The experimental values that are reported in Fig.3a in units of $E_c$, on the contrary, are significantly reduced from $\Delta_{SAS}(\theta)/E_c$ (values reported on the horizontal scale) due to the impact of inter-layer electron correlation. Such correlation effects that yield a softening of the mode can be captured by a simple expression that links the splitting to the density of excitonic pairs $n_{ex}$ (values are reported in the supplementary material) through the relation[7]:

$$\delta(\theta) = \Delta_{SAS}(\theta) \bullet [(n_S - n_{AS})/n_T] = \Delta_{SAS}(\theta) \bullet [1 - 2n_{ex}/n_T].$$

The correlated excitonic state can be also described as an imperfect pseudospin ferromagnet (pseudospin $\tau_x = \pm 1$ describes occupation of symmetric/antisymmetric Landau levels) with a pseudospin polarization (or order parameter) $<\tau_x> < 1$. Within this framework $\delta(\theta) = \Delta_{SAS}(\theta) \bullet <\tau_x>$[15], and a continuous second-order QPT is expected when $<\tau_x> = 0$ and the $SF_{SAS}$–SW energy splitting collapses. Current theories interpret the reduction of $<\tau_x>$ in terms of quantum fluctuations of the pseudospin ferromagnet[29].

In contradiction with this picture, the data in Fig.2 and Fig.3a demonstrate a QPT linked to the disappearance of the $SF_{SAS}$ excitation at a finite value of $\delta(\theta)$, which is the correlated gap and an order parameter of the excitonic phase. These striking results are unambiguous evidence of a first-order character of the transformation. Figure 3(a) is revealing. The evolution of $\delta(\theta)$ in the excitonic phase suggests a continuous QPT at an angle slightly above 40°. There is, however, a marked discontinuity in the value of $\delta(\theta)$ at the critical angle ($\theta_c = 34.2°$) that demonstrates the first-order character of the QPT and that highlights the subtle competition of the collective excitonic state with the CF metal phase.

Evidence of the CF phase is found in the spectral lineshape of the SW mode that changes remarkably just above the critical angle as seen in the spectra at $\theta = 34.4°$, $34.6°$ and $35°$ in Figs.1 and 2 (the supplementary material incorporates a more detailed comparison). Here the CF quantum phase is characterized by a low-energy continuum of spin excitations that appears below the SW mode. The continuum of $SF_{CF}$ excitation modes arise from spin-flip transitions across the Fermi energy of the CF metal[10] as indicated in the schematic drawing of CF energy levels shown in the right-middle panel of Fig. 1.

The first-order character of the QPT is further seen in the abrupt collapse of the $SF_{SAS}$ integrated intensities at angles above $\theta_c = 34.2^o$ that is shown in Fig. 3b. The collapse links directly with the disappearing of the excitonic phase as the critical angle is approached. We note, in addition, that the precision achieved in determining the value of the critical angle reflects the electron density (i.e. filling factor) uniformity within the quantum Hall excitonic phase. This stems from the observed large dependence of $\theta_c$ on filling factor values (see supplementary material). Taking this into account we can conclude that filling factor fluctuations ($\delta\nu_T$) around $\nu_T=1$ within the excitonic phase are smaller than $=1.2 \times 10^{-4}$.

The appearance of the $SF_{CF}$ continuum of the CF metal is concomitant with the vanishing of the $SF_{SAS}$ intensity and the collapse of the correlated gap of the excitonic phase to its lowest value as displayed in Figs. 2, 3 (see also supplementary material). This further highlights that the removal of criticality and the consequent weakly first-order character of the transition is truly determined by the competition and interplay between inter-layer and intra-layer correlations. The role of disorder in smearing the sharp QPT is manifested in the temperature dependence of the spectra shown in Fig.4. It can be seen that the 'critical' temperature of the disappearance of the $SF_{SAS}$ mode decreases continuously as the critical angle is approached. This behavior, which is in agreement with a recent analysis of transport data performed in the limit of vanishing tunneling as a function of $d/l_B$[27], suggests that phase coexistence at finite temperatures could be due to competition of the free energies of the two phases. Figure 4 highlights that the collapsed value of the correlated gap $\delta(\theta_c)$ is not affected by temperature in the range where the $SF_{SAS}$ is visible ruling out that its measured finite value at the critical angle is due to the finite temperature of the system.

The topic of the impact of competing correlated phases on quantum phase transitions is the focus of considerable interest in the framework of studies of QH liquids[14,30], and some theoretical works have described competing interaction terms by gauge fields. In this context, there is the intriguing scenario that fluctuations of the gauge fields could lead to competing order parameters. First-order transitions in the absence of disorder are predicted in quantum Hall insulators[14] in other condensed matter systems[17] and in particle physics[16]. One key property of the electron bilayer system studied here is that the roles of different electron correlation terms at the QPT can be tested by finely tuning their relative strength.

## ACKNOWLEDGEMENTS

This work was supported by the projects MIUR-FIRB No. RBIN04EY74, A. P. is supported by the National Science Foundation under Grant No. DMR-0352738, by the Department of Energy under Grant No. DE-AIO2-04ER46133, and by a research grant from the W. M. Keck Foundation. We acknowledge useful discussion with Rosario Fazio.

## METHODS

Spin excitations by resonant inelastic light scattering are measured in a backscattering configuration using a single-mode tunable Ti-Shappire laser at around 810 nm. Laser power densities was kept at $10^{-3} - 10^{-4}$ W/cm$^2$ and a crossed polarization condition, where incident and scattered photons have opposite polarizations, was used. A triple-grating spectrometer in additive configuration equipped with master gratings and a nitrogen-cooled CCD was used for detection. Selection rules for spin excitations in zinc-blend semiconductors such as *GaAs* dictate further that the incident or scattered photons must have a component of the polarization along the growth axis *z,* which is made possible by tilting the sample with respect to the magnetic field direction. Spin-flip processes are accessed by this method thanks to spin-orbit coupling of valence-band states that mixes different spin states. The measurements are performed on the sample mounted on a mechanical rotator in a dilution refrigerator with base temperature ~ 50 mK. Standard Shubnikov-de-Haas measurements were carried out to carefully calibrate the rotational stage. The sample studied in this work is a nominally symmetric modulation-doped $Al_{0.1}Ga_{0.9}As/GaAs$ double quantum wells grown by molecular beam epitaxy with total electron density of $n_T$ ~ 1.1 x $10^{11}$ cm$^{-2}$, mobility above $10^6$ cm$^2$/Vs and measured tunneling gap at zero magnetic field of $\Delta_{SAS}$ = 0.36 meV. Measurements of spin excitations across the Zeeman gap (spin wave - SW) as a function of total magnetic field yields a value for the gyromagnetic factor of $g$=-0.4.

## Captions

Fig.1: **Competing phases in bilayers**. Upper panel: Schematic representation of the correlated phases in double layers. The excitonic quantum Hall phase is shown on the left. Electron-hole pairs occur between symmetric and anti-symmetric spin-up Landau levels. The composite fermion (CF) phase is shown on the right. The Fermi energy indicates the occupied CF levels. Middle panel: Schematic energy level diagrams and spin excitations in the two phases. SW refers to the spin-wave excitation across the Zeeman gap. $SF_{SAS}$ represents the spin-flip mode from spin up symmetric (S) state to anti-symmetric (A) spin down state across the tunneling gap. In the CF phase, a spin flip ($SF_{CF}$) continuum of excitations across the Fermi level extends from the Zeeman gap down to an energy value determined by the relative position of the Fermi level within the spin-up and spin-down CF states. Lower panel: Representative spin excitation spectra after background subtraction due to magneto-luminescence and stray light in the excitonic quantum Hall (left) and CF metallic (right) phases measured by resonant inelastic light scattering at two different tilt angles and $T$= 50 mK. A lorentzian fit to the SW in the CF phase (black line) is shown to highlight the impact of the $SF_{CF}$ continuum (shaded in green).

Fig.2: **Spin excitations of the two correlated phases.** Angular dependence of spin excitations around the critical angle of $\theta_c=34.2^o$. The spectra show the dramatic quench of the spin-flip $SF_{SAS}$ excitation intensity in the excitonic phase (shaded in orange) and the abrupt increase of the spin-flip continuum $SF_{CF}$ (shaded in green) in the CF phase as the tunnelling gap is reduced by finely tuning the tilt angle. Best fit results to the spectra (dotted lines) with two lorentzians (one in the CF phase) and the magneto-luminescence continuum and laser stray light (solid grey lines) are shown.

Fig.3: **Experimental manifestation of the first-order quantum phase transition.** (**a**) Evolution of the correlated gap of the excitonic phase, i.e. the energy splitting between SW and $SF_{SAS}$ excitations, as a function of tunnelling gap or tilt angle. The splitting remains finite at the phase boundary between the excitonic phase (red shaded) and the composite fermion metal (blue shaded) preventing the continuous phase transition expected at the collapse of the gap shown as dashed line. (**b**) Plot of integrated spin-flip excitation intensity. Data corresponds to the $SF_{SAS}$ (white) and $SF_{CF}$ (yellow) intensity (solid and dotted lines are guide for the eyes). The values are obtained after normalization with respect to the spin-wave SW intensity to take into account the angular dependence of oscillator strength of spin excitations.

Fig.4: **Impact of temperature.** (**a,d,g**) Temperature dependence of spin excitations in the excitonic phase at filling factor $v_T = 1$ for three different tilt angles. (**b,e,h**) Normalized spin-flip $SF_{SAS}$ intensity as a function of temperature and tilt angles. (**c,f,i**) Plot of SW-$SF_{SAS}$ energy splitting versus temperature. The correlated gap of the excitonic phase remains constant within uncertainty as temperature increases.

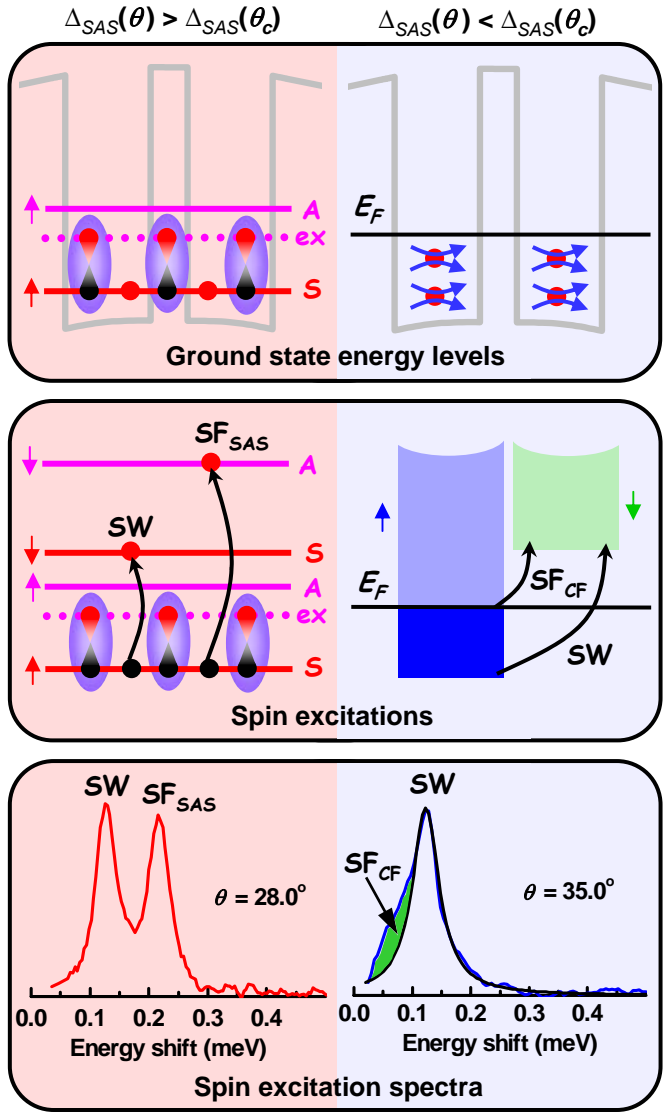

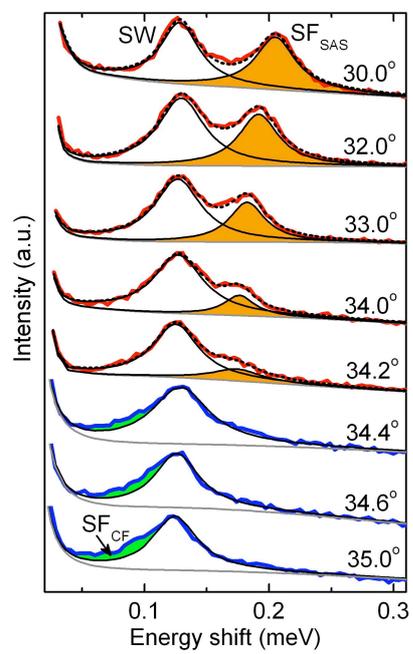

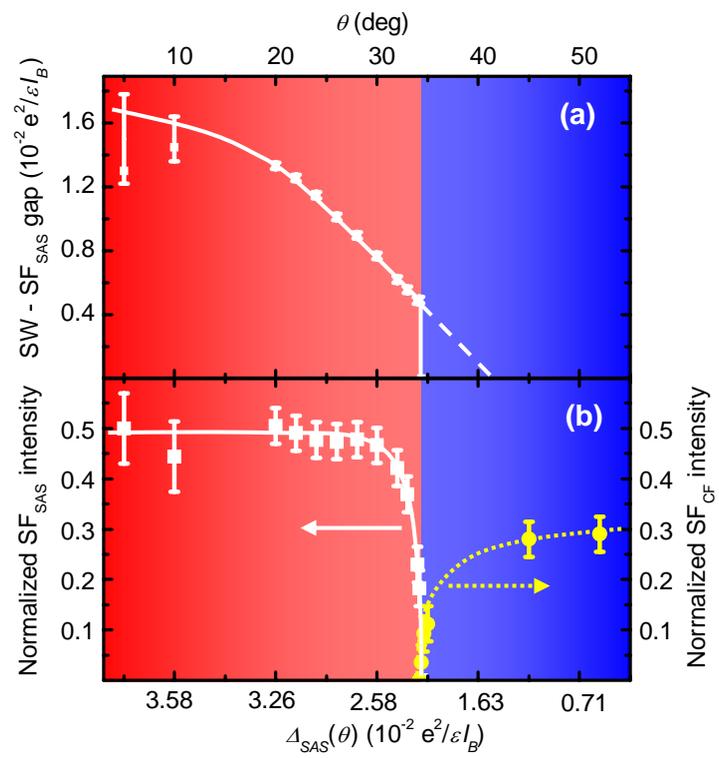

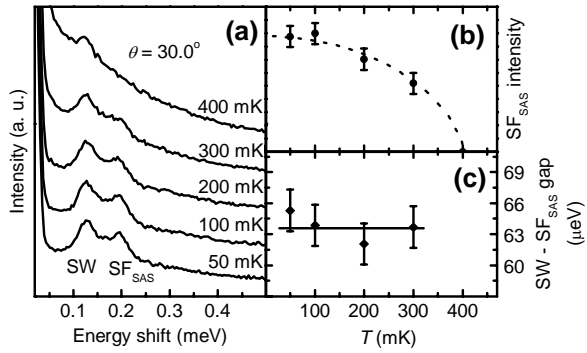
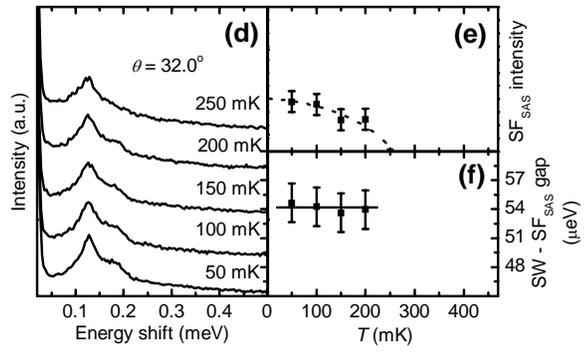
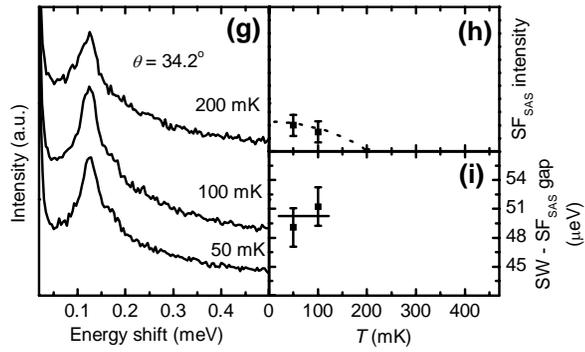

# Supplementary Materials

**Experimental resolution of phase separation:** Figure 2 in the main text of the paper highlights that the $SF_{SAS}$ excitation is suddenly replaced by the $SF_{CF}$ continuum with an increase of tilt angle of 0.2° at $\theta_c$=34.2°. To further address this abrupt modification we compare in Fig.1supp. two spectra acquired at $T$=50 mK. The changes of line shapes, with disappearance of the $SF_{SAS}$ mode (orange) and appearance of the $SF_{CF}$ continuum (green), occurs at the tunneling gap critical value of $\Delta_{SAS}/E_C = 2.21 \times 10^{-2} \pm 1 \times 10^{-4}$.

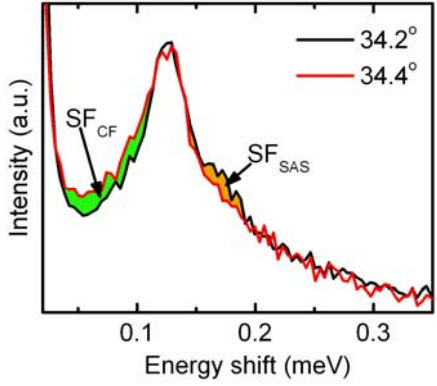

**Figure 1supp:** Comparison of two spectra at the phase transition. $SF_{SAS}$ and $SF_{CF}$ excitations are highlighted with orange and green colors respectively.

**Filling factor uniformity of the excitonic quantum Hall phase:** Electron density (i.e. filling factor) and tunneling gap spatial non-uniformities due to residual disorder of the two-dimensional electron gas in the incompressible phase can change dramatically the manifestation of the quantum phase transition. One effect of these fluctuations is to create compressible islands in the incompressible quantum Hall phase. This phase co-existence manifests in resonant Rayleigh scattering[30] in the quantum Hall phase and in other magneto-transport experiments with peculiar temperature-dependent behavior[29,31]. Our results, however, support the surprising conclusion that the incompressible excitonic regions are very uniform. This uniformity is crucial for the observation of the first-order character of the quantum phase transition. For a quantitative estimation of the uniformity of the excitonic quantum Hall phase, we studied the evolution of critical angle $\theta_c$ with filling factor (Fig.2supp). It can be seen that the critical angle decreases markedly by small changes in filling factors. From these data and considering our experimental resolution of the critical angle of $\delta\theta_c = 0.2°$ we obtain that the average filling factor fluctuation in the excitonic phase must be smaller than $\delta\nu_T \sim 1.2 \times 10^{-4}$. This uniformity of the incompressible excitonic phase and the possibility to probe selectively the excitonic regions by means of the $SF_{SAS}$ excitation makes it possible the observation of the discontinuous quantum phase transformation with resolution in the tunneling gap values of of $\delta(\Delta_{SAS}/E_C) = 2 \times 10^{-4}$.

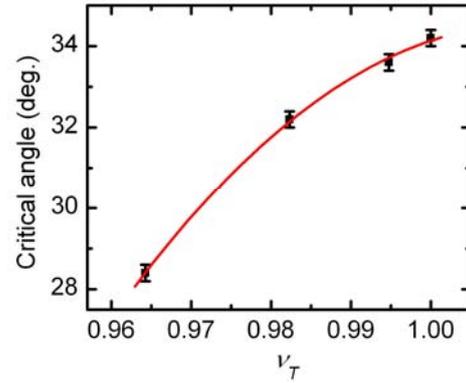

**Figure 2supp:** Critical angle versus filling factor $\nu_T$. The red curve represents a polynomial fit to the data. A similar decrease of the critical angle is observed with increasing filling factor $\nu_T > 1$. In our analysis, the average slope $\delta\theta_c/\delta\nu_T$ of the data points is used to estimate the degree of spatial uniformity of filling factor in the excitonic quantum Hall phase.

**Exciton density and order parameter:** The measured energy splitting between SW and $SF_{SAS}$ excitations is smaller than the bare tunneling gap $\Delta_{SAS}$ as shown in Fig.3a in the main text. This effect is determined by the reduction of the degree of pseudospin polarization in the quantum Hall state[15]. This is a consequence of the formation of excitons across the tunneling gap due to interlayer correlations. The pseudospin order parameter $\langle\tau_x\rangle$ is directly linked to the exciton density $n_{ex}$ as described in eq.1. The evolution of the exciton density versus tunneling gap reported in Fig.3supp shows an enhancement of the exciton density as the phase boundary is approached. At the phase boundary the exciton density does not reach 50%, which corresponds to full depolarization of the pseudospin ($<\tau_x> = 0$). The finite pseudospin polarization at the phase boundary is regarded as a signature of the first-order character of the quantum phase transition at total filling factor $\nu_T = 1$.

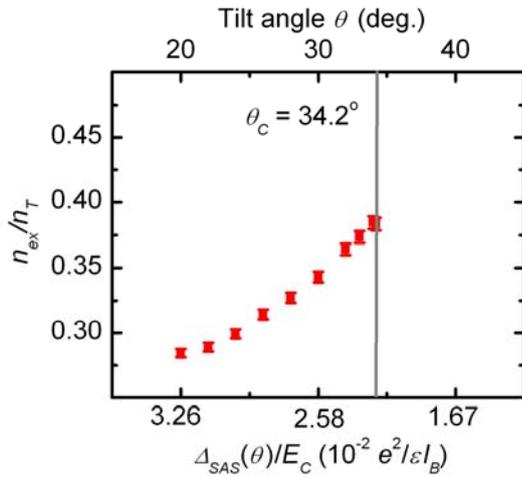

**Figure 3supp:** Evolution of exciton density $n_{ex}$ normalized to the total electron density $n_T$ versus tunneling gap and tilt angle.